\documentclass[11pt]{article}
\usepackage{multicol}
\usepackage{graphicx}
\usepackage{amssymb,bm,mathrsfs,bbm,amscd}
\usepackage[tbtags]{amsmath}
\usepackage{lastpage}

\begin{document}

\title{ LWFA as a Preinjector for XFEL Driver Linac }
\author{ Xiongwei Zhu \\
{Institute of High Energy Physics, Chinese Academy }\\
 {of Sciences, P.O.Box 918, Beijing 100049, China}}

\maketitle

\begin{abstract}
   In this paper, we propose to use the LWFA as the preinjector for XFEL driver linac. We can use LWFA to produce the femtosecond
electron beam. The peak current of the produced beam can reach tens of $kA$ and is high enough to drive XFEL so that the present bunch compressing technique can be deserted. The output optical pusle length can be as short as $ 1 fs $.

\end{abstract}


\section{ Introduction }
\qquad LWFA is one of the cadidates for the future accelerator\cite{1,2}. On the high energy frontier, people want to use LWFA technique to construct compact linear collider; On the high quality beam frontier, the researchers try to use LWFA as FEL driver. There are many kinds of conventional electron guns, such as high voltage DC electron gun, microwave electron gun. Plasma based
electron gun is a new kind of electron gun. Free electron laser is the full coherent light source and needs high brightness electron injector.
Low emittance and high peak current are the aim we pursue. The conventional gun has the limit. Up to now, the best achieved normalized emittance of the photoinjector is $1 mm mrad$ with the bunch charge of $1 nC$. If we need more bright electron bunch, we may use the new mechanism to produce the high quality electron beam.

Laser plasma based accelerator is a key research highlight in the field of new concept accelerator\cite{3,4,5}. In the past, there are some proposals\cite{6,7,8,9} to use LWFA as the FEL driver linac directly. But it is difficult to get the stable high energy
electron beam with the present LWFA experiment. So, LWFA may be the potential electron linac preinjector. This new kind of electron gun has the advantages of the ultrashort bunch length, the ultra-low emittance, and the high peak current quasi-monoenergetic electron beam. It is unnecessary to use the bunch compressor to get the high peak current beam with our proposed electron preinjector.

\section{ Laser wakefield accelerator }
\qquad Laser wakefield accelerator ( LWFA )\cite{2,3,4,5} had a great breakthrough during the past 4-5 years. With the discovery of the bubble mechanism,
the 1 GeV or so quasi-monoenergetic electron beam is obtained in the LWFA experiments\cite{3,4,5}. The energy spread is still several percent and can not used as the FEL driver. But this produced beam can be used as the high brighness electron injector due to their low normalized emittance, ultrashort bunch length, and high peak current. In this case, we just need tens of MeV high quality electron bunch to be used as the preinjector of the driver linac.

\section{ Application for next generation light source-Free Electron Laser }
\qquad In\cite{6}, we propose a new mechanism LWFA to produce femtosecond electron bunch with the bunch length of $ 2 fs $, and the peak current of $ 750 A $. this kind of electron bunch can be used to produce ultrashort Terahertz coherent radiation. The peak current is also
high enough to drive soft X-ray free electron laser, if we boost the beam energy to $ 1 GeV $ or so.

In this paper, as an example, we give the 2D PIC simulation\cite{10} example result of one LWFA experiment design with the peak current in the range of $ 15-20 kA $. The used code is Vorpal\cite{10}, Vorpal is an object-oriented PIC code developed by Tech-X Corporation. The typical laser paremeters are the wavelength of 800 nm, the pulse length of $ 1.6 \mu m $, the beam waist of $ 2.2 \mu m $ and the peak intensity of $ 1.63 \times10^{20} W/cm^2 $. The typical plasma parameters are the plasma density of $ 5.99 \times10^{19} /cm^3 $. Table 1 gives the main parameters.
\begin{center}
{Table 1.Typical LWFA parameters}

\begin{tabular}[width=10cm]{@{\extracolsep{\fill}}|c|c|}
\hline
\multicolumn{2}{|c|}{ Laser}\\
\hline
Wavelength & $ 800 nm$   \\
\hline
Waist & $2.2 \mu m$  \\
\hline
Rep & $10 Hz$ \\
\hline
Peak Intensity & $ 1.63\times10^{20} W/cm^2 $  \\
\hline
Pulse length & $ 5.3 fs$ \\
\hline
\multicolumn{2}{|c|}{ Plasma}\\
\hline
Density & $ 5.99\times10^{25} m^{-3} $\\
\hline
Plasma wavelength & $ 4.32 \mu m $\\
\hline
\multicolumn{2}{|c|}{ Output beam }\\
\hline
Energy & $ 37 MeV$   \\
\hline
Normalized emittance & $1.2 \mu m$  \\
\hline
Rep & $10 Hz$ \\
\hline
Energy spread & $ 3.9\% $  \\
\hline
Bunch charge & $ 40 pC $ \\
\hline
\end{tabular}

\end{center}

Under these parameters, we can get the electron beam with the energy
of $ 37 MeV $, the bunch charge of $ \sim 40 pC $, the bunch length of $ \sim 1 fs $, the normalized emittance of $ 1.2 mm mrad $, and the rms energy spread of $ 3.9 \% $. Figure 1 shows the output 2D density distribution of the produced electron beam.

\begin{center}
\includegraphics[width=10 cm]{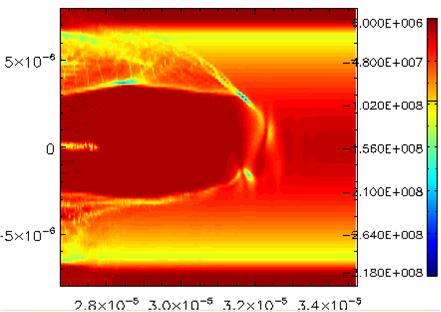}

{ Figure 1.The 2D electron density (the colour stands for the density strength).}
\end{center}

We analyze the phase space of the bunched beam in the bubble. Figure 2 shows the output beam phase space.
\begin{center}
\includegraphics[width=10 cm]{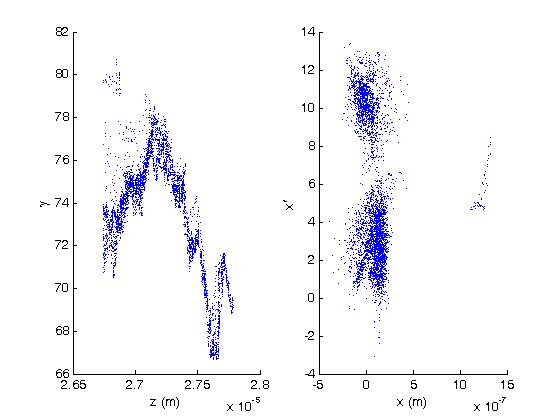}

{ Figure 2.The longitudinal and transverse phase space. }
\end{center}
Figure 3 shows the current profile, the peak current can reach $ 15-20 kA $.

\begin{center}
\includegraphics[width=10 cm]{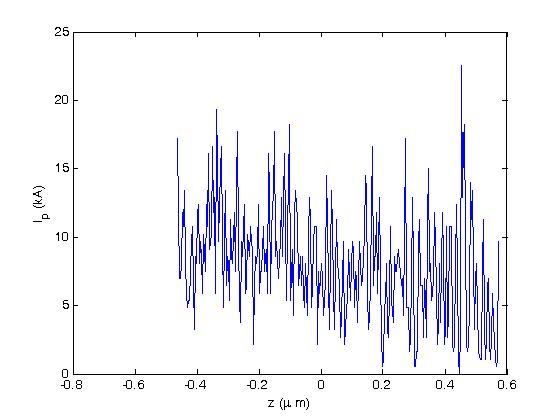}

{ Figure 3.The current profile. }
\end{center}

Then, we use this injector to set up soft x-ray free electron laser experiment. We add the booster linac after the lwfa injector to boost the beam energy to the energy of 1 GeV. The linac can choose SLAC $2856 MHz$ S-band accelerating structure or Spring-8 $5712 MHz$ accelerating structure. Due to the high peak current and femtosecond bunch length, we don't use the bunch compressor to
obtain the high peak current bunch to drive XFEL process. We use the typical beam parameters: the energy of $1GeV$, the normalized emittance of $ 1.2 mm mrad $, the energy spread of $ 0.1\% $, and the peak current of $ 15-20 kA $.

\begin{center}
\includegraphics[width=10 cm]{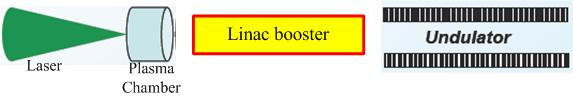}

{ Figure 4.The schematic layout of XFEL experiment. }
\end{center}

We use M.Xie's fitting formula\cite{11} to estimate the SASE mode FEL output. Using the mature undulator parameters: the period of $ 3 cm $, and the gap of $ 1 cm $. The typical output parameters are the wavelength of $ 12.5 nm $, the peak output laser power of $ 90 GW $, the saturation length of $ 8 m $, and the gain length of $ 0.33 m $. The analytically estimated peak power is bigger than the simulated peak power ($\sim 40 GW$) by GENESIS. The main parameters are summaried in Table 2.
\begin{center}
{Table 2.Main parameters of the soft X-ray}

\begin{tabular}[width=10cm]{@{\extracolsep{\fill}}|c|c|}
\hline
\multicolumn{2}{|c|}{ Electron beam }\\
\hline
Energy & $ 1 GeV$   \\
\hline
Normalized emittance & $1.2 \mu m$  \\
\hline
Rep & $10 Hz$ \\
\hline
Energy spread & $ 0.1\% $  \\
\hline
Bunch charge & $ 40 pC $ \\
\hline
Peak current & $15-20 kA$ \\
\hline
\multicolumn{2}{|c|}{ Undulator }\\
\hline
Period & $ 3 cm $\\
\hline
Gap & $ 1 cm $\\
\hline
K & $  $\\
\hline
Saturation length & $ 8 m $ \\
\hline
\multicolumn{2}{|c|}{ Photon }\\
\hline
Wavelength & $ 12.5 nm $\\
\hline
Peak power & $40 GW $ \\
\hline
Pulse length & $ 1 fs $\\
\hline
Rep & $ 10 Hz $ \\
\hline
\end{tabular}

\end{center}
Using GENESIS\cite{12} to do time independent and dependent simulation, the typical output peak power is shown as
\begin{center}
\includegraphics[width=10 cm]{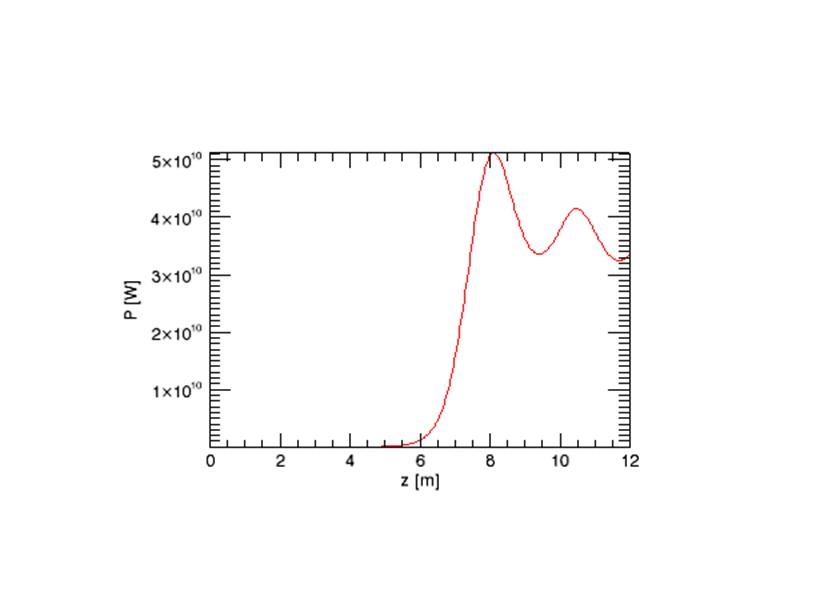}

{ Figure 5.The output peak power along the distance. }
\end{center}
the simulation shows the output peak power is about $ 40 GW $ which is smaller than the analytically estimated value , and the saturation length is about $ 8m $. Figure 6 shows the output laser spectrum.

\begin{center}
\includegraphics[width=10 cm]{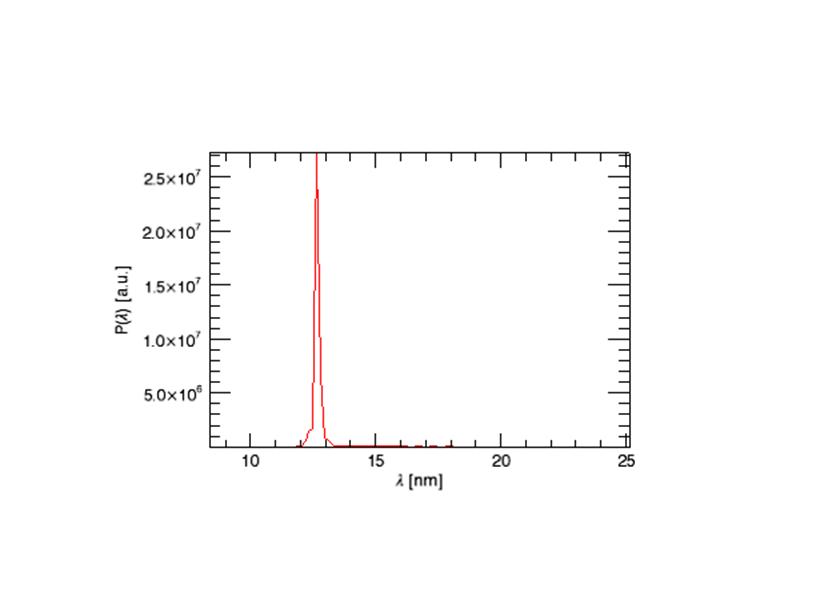}

{ Figure 6.The output laser spectrum. }
\end{center}

\section{ Discussion }
\qquad In order to use LWFA as the electron injector, we have to solve some issues, such as the stability problem, the timing problem, etc.
As we know, the conventional electron injector is more stable than LWFA. As the output electron bunch is in the order of femtosecond, the timing jitter will also be in this order. Therefore, the timing jitter problem maybe not a serious problem for our case.

\section*{Acknowledgement}
\qquad This work is supported by the Innovation Fund of Chinese Academy of Sciences.

\end{document}